\documentstyle[12pt,epsfig]{article}
\begin{document}
\baselineskip=15pt
\begin{center}
{\bf The appearence of the resolved singular hypersurface
${x_0}{x_1}-{{x_2}^n}=0 $ in the classical phase space of the 
Lie group $SU(n)$} 
\end{center}
\begin{center}
{\bf Samir K. Paul}\footnote{smr@bose.res.in} \\
S. N. Bose National Centre For Basic Sciences
Block-JD, Sector-III, Salt Lake\\
Calcutta-700098,  India
\end{center}
\begin{center}
{\bf Siddhartha Sen}\footnote{sen@maths.tcd.ie}\\
School of Mathematics,Trinity College
Dublin,Ireland\\
and\\
IACS , Jadavpur , Calcutta-700032 , India\\
\end{center}
\vspace{1.0cm}

\begin{abstract} 
A classical phase space with a suitable symplectic structure is constructed 
together with functions which have Poisson brackets algebraically identical to 
the Lie algebra structure of the Lie group $SU(n)$. In this phase space we 
show that the orbit of the  generators corresponding to the simple roots of 
the Lie algebra give rise to fibres that are complex lines containing spheres. 
There are $n-1$ spheres on a fibre and they intersect in exactly the same way 
as the Cartan matrix of the Lie algebra. This classical phase space bundle,
being compact, has a description as a variety. Our construction shows that 
the variety containing the intersecting spheres is exactly the one obtained 
by resolving the singularities of the variety ${x_0}{x_1}-{{x_2}^n} = 0$
in ${\mathcal C}^3$ . A direct connection between this singular variety 
and the classical phase space corresponding to the Lie group $SU(n)$ is 
thus established.
     
\vspace{0.5cm}

PACS NO.:  02.40.Xx , 02.20.Bb , 02.40.-k , 11.25.-w

\end{abstract}

\vspace{2.00cm}
\newpage 

\noindent        {\bf I~~Introduction}

\vspace{0.2cm}
 It has long been known that there is an intriguing algebraic correspondence 
between the Cartan matrix of simply laced Lie groups and the intersection 
matrix of spheres that appear when certain simple singularities are
resolved$[1]$. 
The reason for such a correspondence has also been long known within the 
framework of algebraic groups [2] . That this correspondence might be more 
than a mathematical curiosity was established when it was shown that 
duality in string theory made effective use of such a link$[3]$. A type 
2A string compactified on a 
$K3$ surface (a four dimensional surface) was conjectured to be dual to a 
heterotic string compactified on $T^4$ (the four torus). A test of this 
conjecture
required the zero mass excitations in the two theories to match.The zero mass 
excitations at the type 2A end came from certain singular points that appear 
on the $K3$ surface in a certain limit while those at the heterotic string end
came from gauge excitations associated with an $SU(n)$ Lie group.The excitations at
 the type 2A end were "classical" solitonic type excitations while
 those at the heterotic end were "quantum" gauge excitations. This result 
 suggests that a link between minimally resolved singularities and the 
 "classical limit" of simply laced Lie groups might exist.\\

In this paper we establish such a link. We demonstrate this link 
explicitly for the Lie groups $SU(2)$ , $SU(3)$ and $SU(4)$ . Generalisation to 
$SU(n)$ is then straightforward. We find that using a Gauss decomposition   $Z_+ H 
Z_-$ , where $Z_+$ is an  upper triangular matrix with unit diagonal elements ,
$H$ is a diagonal matrix, $Z_-$ is a lower triangular matrix with unit diagonal elements that the    classical phase space, 
in which Poisson brackets mirror the Lie algebra structure, can be constructed 
using a standard  coherent state approach [4].We use the Gauss decomposition in
order to use the Borel-Weil Theorem. This Theorem shows how irreducilbe
representations of a compact Lie group G can be constructed as holomorphic
sections over $G/T$ where T is the maximal torus of G. The approach
thus describes the group in terms of complex variables and is thus a
natural setting for making contact between the group and complex algebraic 
varieties. We have\\ 
${\bf Borel-Weil }$  ${\bf Theorem~~1 }~[5]$ 
$The$ $space$ $of$ $holomorphic$ $sections$ $of$ $the$ $line$ $bundle$ 
$L_\lambda$ $over$ $G/T$ $is$ $non$ $trivial$ $if$ $\lambda$ $is$ $the$ $highest$ $weight$
$of$ $an$ $irreducible$ $representation$
$of$ $G$. $When$ $\lambda$ $is$ $the$ $highest$ $weight$ $then$ $this$ $space$ $of$ $holomorphic$
$sections$ $is$ $a$ $realisation$ $of$ $the$ $representation$ $space$ $V_\lambda$ $of$ $G$.\\ 
The coherent state approach is an explicit way of implementing the Borel-Weil 
Theorem. Use is made, in this approach, of the fact that $G/T$=$G_c/B$ where 
$G_c$ is the complexification of $G$ and
$B$  is the Borel subgroup. In terms of the Gauss decomposition $B$ is generated
by $H$ and $Z_-$. It is in this framework that we search for and identify
classical phase space. For $SU(n)$ the phase space is $CP^{n-1}$. This is
a K\"ahler symplectic manifold. It is identified cleanly in $G_c/B$ as follows..
We first introduce some definitions.  Triangular matrices with unit 
diagonal elements are called unipotent. $Z_+$,$Z_-$ are unipotent.We claim that
by using  
unipotent group elements  
 of a special kind 
we can  construct the required K\"ahler  symplectic manifold.  To do this 
the unipotent elements used must have two  special features. 
 First  they must  mutually
commute . Second the number of elements that commute
with the these unipotent elements must equal  the rank $r$ of the group. Such 
elements are called regular.In this coherent 
state approach the symplectic structure is derived from a K\"ahler potential
which we construct from a scalar product using the group elements described
acting on the highest weight vector for the fundamental representation. It is     then shown  that
 the expectation value of the generators 
of the Lie algebra have Poisson brackets, defined in terms of this 
symplectic structure,isomorphic to the Lie algebra 
of the group. On this K\"ahler phase space a fibrebundle structure can be           constructed whose  
fibres are complex curves, containing intersecting
spheres. The fibre bundle considered  is
constructed from the orbit of group elements arising from the simple roots
of the corresponding Lie algebra.In our construction we classify the unipotent  elements of
$Z_+$  in the following way :\\
One class have centralizer (elements of the group commuting among themselves) 
of dimension equal to the rank $r$ of the group. These are the regular elements. The other class comes from 
the generators of the Lie algebra for simple roots. These are  unipotent         elements that commute with $r+2$ group elements . Such elements are called 
subregular .
 These unipotent subregular elements play a crucial role
in our construction. The special unipotent regular elements are used
to construct
 classical phase space. This is done by acting on the highest weight vector
 with the the elements described. Choosing to work with the highest weight
vector is the way 
the quotienting of $G_c$ by $B$ is implemented in this framework.
The unipotent subregular elements  acting on a point of this  phase space  
 give rise to a fibre containing  intersecting spheres. The way these
 spheres intersect can be summarised in the form of an intersection matrix.
 This intersection matrix is found to be identical to the negative of the Cartan
 matrix of the Lie algebra. We 
show this explicitly  for $SU(3)$ where there are  two  intersecting 
spheres  , for $SU(4)$ where there are three spheres and for $SU(n)$ where      there are   $n-1$ spheres.
Precisely such intersecting spheres    also appear when certain singular     points 
 are resolved on the hypersurface ${x
_0}{x_1}-{x_2^n}=0$ where $({x_0} , {x_1} , {x_2})\in{{\mathcal C}^3}$ , the 
space of three complex variables . 
 In   section-II we summarise the basic facts we need from the theory of 
resolution of singularities. In section-III the classical phase space for the    Lie groups  
 $SU(2)$ , $SU(3)$ and $SU(4)$ is constructed and the link with the resolved
 singularities established. Finally in  section-IV we    summarise our 
 conclusions.  
\vspace{0.5cm}

\noindent       {\bf II~~ Resolving Singularities}

\vspace{0.2cm}

 Consider the hypersurface $V_n$ in ${{\mathcal C}^3}$ defined by the algebraic
 equation:
\begin{equation}
{V_n}({x_0},{x_1},{x_2}) = {x_0}{x_1} - {x_2^n} = 0
\end{equation}
 where $n$ is an integer $\ge 2$ and $({x_0},{x_1},{x_2})\in{{\mathcal C}^3}$. 
We have the following definition$[1,6]$ :\\
{\bf Definition} : $A$ $point$ $({x_0},{x_1},{x_2})\in {{V_n}=0}$ $is$ $a$ $singular$ 
$point$ $of$ $the$ $hypersurface$ $if$ ${\partial_{x_i}}{V_n} = 0$ $at$ $that$ $point$.\\
 It follows from the definition that the point $(0,0,0)$ i.e. the origin is a 
 singular point of the hypersurface ${V_n}\equiv {x_0}{x_1} - {x_2^n}=0$ , for $n
\ge 2$. Indeed a simple definition of this hypersurface as an orbifold is 
possible. To see this set ${x_0}={{\xi}^n}$, ${x_1}={{\eta}^n}$, ${x_2
}={\xi}{\eta}$
 where $({\xi},{\eta})\in {{\mathcal C}^2}$. We note that in terms of these 
 variables ${\xi}, {\eta}$ the equation ${x_0}{x_1} - {x_2^n} = 0$ is 
 identically satisfied i.e. ${\xi} , {\eta}$ parametrise the hypersurface ${V_n}=0$. There is
however one restriction on the variables ${\xi}, {\eta}$ when they are on the
 hypersurface ${V_n}=0$ namely the point $({\xi}, {\eta})$ must be identified  	
with $({\omega^{\frac{1}{n}}}{\xi}, {\omega^{\frac{1}{n}}}{\eta})$ where $\omega$ is an $nth$ root of 
identity $({\omega^n}=1)$. Thus the hypersurface ${V_n}=0$ can be 
identified with the orbifold ${{\mathcal C}^2}/{{\mathcal Z}_n}$ 
with ${\mathcal Z}_n$ action defined by 
$({\xi},{\eta})\longrightarrow ({\omega^{\frac{1}{n}}}{\xi},
{\omega^{\frac{1}{n}}}{\eta})$, ${{\omega}^n}=1 $.\\
There is a standard method of minimally resolving this singularity [1,6] i.
e. of constructing a globally well defined hypersurface which is in $1-1$ 
correspondence with the original hypersurface ${V_n}=0$ except at the point $(0,0,0)$. 
The singular point is "blown up" . We describe this procedure first for the case $n=2$ and $n=3$ and then for the general case where $n>3$.\\
 Let us introduce the space ${{\mathcal C}^3}\times {{\mathcal P}^2}$ where       ${{\mathcal P}^2}$
 is the complex projective two space (henceforth we denote 
${\mathcal CP}^n$ by ${\mathcal P}^n$ ). Points in ${{\mathcal C}^3}\times
{{\mathcal
 P}^2}$ can be written as the pair $( ({x_0},{x_1},{x_2}) , [{s_0},{s_1},{s_2}]
 )$ where $({x_0},{x_1},{x_2})\in {{\mathcal C}^3}$ and $[{s_0},{s_1},{s_2}]$ i
s an element of ${{\mathcal P}^2}$ i.e. it represents the equivalence class of 
points $({s_0},{s_1},{s_2})$ under the equivalence relation 
$({s_0},{s_1},{s_2})\sim {\lambda}({s_0},{s_1},{s_2})$ where $\lambda$ is a complex 
number $\ne 0$. Next we introduce the space ${{\mathcal C}^3}( {{\mathcal 
P}^2},{\mathcal R} )$. This is defined as the set:
\begin{equation}
{{\mathcal C}^3}({{\mathcal P}^2},{\mathcal R})=\{ ({x_0},{x_1},{x_2}),[{s_0},{
s_1},{s_2}]\vert {x_i}{s_j}={x_j}{s_i} , \forall i,j \}
\end{equation}
Geometrically the restriction ${x_i}{s_j}={x_j}{s_i}$ means that $s_i$ is
proportional to $x_i$. This
gives a space consisting 
of points $({x_0},{x_1},{x_2})$ in ${\mathcal C}^3$ and lines through the 
origin and these points. These lines are elements of ${{\mathcal P}^2}$. Thus for all
points in ${{\mathcal C}^3}$ , other than the origin , the element of
${{\mathcal P}^2}$ is uniquely fixed by $({x_0},{x_1},{x_2})$. There is 
thus an $1-1$ correspondence between points in ${{\mathcal C}^3}$ and the pair 
of points in 
${{\mathcal C}^3}( {{\mathcal P}^2},{\mathcal R} )$ defined by the eq.(2).
For the origin however the 
situation is different. When ${x_0}={x_1}={x_2}=0$ , there is no restriction on
$[{s_0},{s_1},{s_2}]$. Thus the origin of ${{\mathcal C}^3}$ is replaced by the
 entire ${{\mathcal P}^2}$ in ${{\mathcal C}^3}( {{\mathcal P}^2},{\mathcal R} 
)$
 ; it is "blown up". Let us now study the way the hypersurface ${x_0}{x_1}-{x_2
^2}=0$ behaves in ${{\mathcal C}^3}({{\mathcal P}^2},{\mathcal R})$. To see the 
way the singular point in ${V_2}=0$ in ${{\mathcal C}^3}$ gets mapped in 
${{\mathcal C}^3}({{\mathcal P}^2},{\mathcal R})$ we approach the origin in
${\mathcal C}^3$. This is done by scaling  the points $({x_0},{x_1},{x_2})$ in 
${\mathcal C}^3$ by $t$ and letting $t\longrightarrow 0$. Note that the 
constraints ${x_i}{s_j}={x_j}{s_i} \forall i,j$ imply that ${x_i}=k{s_i}$ 
(where $k=constant$). Thus $(t{x_0},t{x_1},t{x_2})=tk({s_0},{s_1},{s_2})$  i.e.,
we get from eqs.(1) and (2)in the $t\longrightarrow 0$ limit , points on 
${V_2}=0$ satisfy ${s_0}{s_1}-{s_2^2}=0$ in ${\mathcal P}^2$. We now have the 
following theorem.\\ {\bf Theorem 2} [7] $A$ $polynomial$ $equation$ $of$ $degree$ $n$ $in$ 
${\mathcal P}^2$ $describes$ $a$ $compact$ $Riemann$ $surface$ $of$ $genus$ $g$ $with$ 
$g={\frac{1}{2}}(n-1)(n-2)$. \\     
 In our case the polynomial equation ${s_0}{s_1}-{s_2^2}=0$ in ${\mathcal P}^2$
 is of  degree 2 . Hence the surface in ${\mathcal P}^2$ is a genus zero surface
 i.e., topologically it is a sphere. Thus the singular point of the hypersurface 
 ${x_0}{x_1}-{x_2^2}=0$ in ${\mathcal C}^3$ is replaced by a sphere in 
 ${{\mathcal C}^3}({{\mathcal P}^2},{\mathcal R})$. The singularity has been resolved 
 by a process of "blowing up"  tuning the singular point into a sphere.\\
 We next consider the case $n=3$. Repeating the procedure for the $n=2$ case we
 find , the points $[{s_0},{s_1},{s_2}]$ satisfying eq.(1) for $n=3$ i.e. ${V_3
}=0$ and eq.(2) in the viscinity of the origin now have to satisfy the polynomial 
equation ${t^2}({s_0}{s_1}-kt{s_2^3})=0$ i.e., the equation ${s_0}{s_1}=0$ in
the $t\longrightarrow 0$ limit. This gives a pair of spheres  ${{\mathcal P}^
1}$'s in ${{\mathcal P}^2}$ (theorem 2 ) corresponding to setting  ${s_0}=0$, $
{s_1}=0$. These two spheres intersect once at the point $(0,0,1)$ in ${\mathcal
 P}^2$.For $n\ge 4$, we again get the equation ${s_0}{s_1}=0$ in  the limit
 $t\longrightarrow 0$ and a pair of spheres. However the intersection of these
 spheres is still a singular point. To see this we choose to describe 
${{\mathcal C}^3}({{\mathcal P}^2},{\mathcal R})$ by first selecting a point in
 ${{\mathcal P}^2}$ , say , $({s_0},{s_1},{s_2})$ with ${s_2}\ne 0$. Choosing 
 this point does not uniquely fix a point in ${{\mathcal C}^3}$ but gives a line 
through the point $({s_0},{s_1},{s_2})$ and the origin in ${{\mathcal C}^3}$. 
Let us set ${s_2}={y_2}$ ${s_0}={y_0}{y_2}$, ${s_1}={y_1}{y_2}$, where $({y_0},{
y_1},{y_2})\in {{\mathcal C}^3}$. Finally set ${y_2}={x_2}$. Then ${x_0}={y_0}{
y_2}$ , ${x_1}={y_1}{y_2}$ and $0 = {x_0}{x_1} - {{x_2}^n} = {{y_2}^2}({y_0}{y_1}
- {{y_2}^{n-2}}) = 0$. By construction ${y_2}\ne 0$. So  ${y_0}{y_1} - {{y_2
}^{n-2}} = 0$. For $n\ge 4$ this hyperplane has a singularity at the origin ${y_0}={y_1}=
{y_2}=0$ where ${s_0}=0$ , ${s_1}=0$. The process of blowing up has 
to be repeated . The spheres produced by this process of blowing up self intersect
in an invariant way.Following a standard procedure it can be shown that the
self intersection of the spheres can be taken to be $-2$  [1].
We summarise the results presented regarding the way the spheres in the 
"resolved singularity" intersect in the form of a matrix. For $n=3$, we have
the intersection matrix $\left(\matrix {-2 & 1\cr 1 & -2}\right)$. For $n=4$
we have , $\left(\matrix {-2 & 1 & 0\cr 1 & -2 &
 1\cr 0 & 1 & -2}\right)$ . This intersection information can be encoded in the
 form of a Dynkin diagram shown in fig.1
\begin{figure}
\centering
\epsfxsize=0.5in\epsfysize=0.06in
\rotatebox{0}{\epsfbox{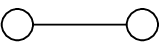}}
\caption{Dynkin diagram 1}
\end{figure}
where each dot denotes a sphere with self intersection -2 and a line joining two
dots intersection between two spheres with intersection number one. The 
construction described here extends to the case of arbitrary $n$ where the 
diagram is shown in fig.2
\begin{figure}
\centering
\epsfxsize=1.5in\epsfysize=0.08in
\rotatebox{0}{\epsfbox{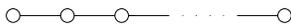}}
\caption{Dynkin diagram 2}
\end{figure}
\vspace{0.3cm}

\noindent {\bf III~~  The classical phase space for the Lie groups $SU(2)$,$SU(3)$
and $SU(4)$} 

\vspace{0.2cm}
We now  look at the classical origins of the Lie groups $SU(2)$,$SU(3)$
and $SU(4)$ in the following sense. The groups have a local structure encoded
by their Lie algebras. We will call the associated phase space , defined with
a suitable symplectic structure , the classical counterpart of the Lie group
if functions on the phase space can be constructed which have Poisson
brackets algebraically identical to the Lie algebra structure of the Lie group. 
The construction we will describe involves coherent states associated with the 
Lie group of interest [4]. We start by quickly summarising the results for $SU
(2)$. This simple example contains a crucial ingredient needed for our subsequent
analysis. Let us introduce the highest weight representation for $SU(2)$, which
we write as the vector $\left(\matrix {0\cr 1}\right)$. The coherent state $
\vert \lambda \rangle$ is then defined as:
\begin{equation}  
\vert \lambda \rangle = {e^{\lambda {J_+}}}\left(\matrix {0\cr 1}\right)
 ,  {J_+} = \left(\matrix {0 & 1\cr 0 & 0}\right) \equiv {e_{12}} 
,  \langle \lambda \vert \lambda \rangle = (1+{\lambda }{\bar \lambda})
\end{equation}
$\lambda$ being a complex variable.  
We then construct the K\"ahler potential\\                                       
   $V({\lambda ,{\bar \lambda}})=k\cdot \log(1+{\lambda {\bar \lambda}})$. This 
gives rise to a symplectic form on the coordinate chart ${\lambda_0}\ne 0$ in
${\mathcal P}^1$ as well as the Fubini-Study metric  [6,8]  where $\lambda ={
\frac{\lambda_1}{\lambda_0}}$ .The symplectic structure is given by 
\begin{equation}
{\omega_{\lambda {\bar{\lambda}}}} = {\partial_\lambda}{\partial_{\bar{\lambda}
}} V(\lambda {\bar{\lambda}}), \omega = {\omega_{\lambda {\bar{\lambda}}}} d\lambda
\wedge d{\bar{\lambda}} + {\omega_{{\bar{\lambda}}{\lambda}}} d{\bar{\lambda}}
\wedge d{\lambda}
\end{equation}
where the symplectic matrix is given by 
\begin{equation} 
[\omega] = k\cdot \left(\matrix {0 & {\frac{1}{{(1+{\lambda {\bar \lambda}})}^2
}}\cr -{\frac{1}{{(1+{\lambda {\bar \lambda}})}^2}} & 0}\right)                
\end{equation}
k is a constant. We next note that
\begin{eqnarray}
{X_+} &=& {\frac{\langle \lambda \vert {J_+} \vert \lambda \rangle}{\langle
\lambda 
\vert \lambda \rangle}} = {\frac{2\lambda}{1+{\lambda {\bar \lambda}}}}\nonumber\\ 
{X_-} &=& {\frac{\langle \lambda \vert {J_-} \vert \lambda \rangle}{\langle \lambda
\vert \lambda \rangle}} = {\frac{2{\bar \lambda}}{1+{\lambda{\bar \lambda}}}
} 
\nonumber\\  
{X_0} &=& {\frac{\langle \lambda \vert {J_0} \vert \lambda \rangle}{\langle \lambda
\vert \lambda \rangle}} = {\frac{1-{\lambda {\bar \lambda}}}{1+{\lambda{\bar
 \lambda}}}}
\end{eqnarray}
with ${J_-} = {e_{21}}$ ,${J_+}={e_{12}}, {J_0} = {e_{11}}-{e_{22}}$,
where  ${e_{ij}}$ stands for the $3\times 3$ matrix with one in the 
${ij}$th position and zero elsewhere, 
and ${X_+}$,${X_-}$,  ${X_0}$ 
are  functions on the phase space ${\mathcal P}^1$ described by the complex
variable $\lambda$ and the symplectic form $\omega$ given by eq.(4).

Furthermore ${X_+}{X_-}+{{X_0}^2}=1$ i.e these functions represent points on
$S^2$.
 Also
\begin{eqnarray}
\{ {X_+},{X_-} \}& = &{{({{\omega}^{-1}})}_{\lambda{\bar \lambda}}}{
\partial_{\lambda}}{X_+}{\partial_{\bar \lambda}}{X_-}+{{({{\omega}^{-1}})}_{
{\bar \lambda
}\lambda}}{\partial_{\bar \lambda}}{X_+}{\partial_{\lambda}}{X_-} \nonumber\\  
   & = &2i{X_0}\nonumber\\
\{ {X_0},{X_{\pm}} \} & = &{\pm}{X_{\pm}}
\end{eqnarray}
for suitable choice of $k$. Thus the expectation values of the generators ${J_{
\pm}}$ , $J_0$ in the normalised state vector $\vert \lambda \rangle$ represent
 the classical functions whose quantisation , achieved by replacing Poisson brackets
 by commutators leads to the Lie algebra structure. The classical phase space
 of $SU(2)$ is thus $S^2$ or ${\mathcal P}^1$ . Note that the presence of 
$S^2$ could be spotted simply by evaluating $\int \omega d\lambda \wedge d{\bar 
\lambda} = 4\pi$ , where $\omega$ is given by eq. (4) and noting that the curvature
of the phase space manifold is constant and positive. Also the metric on the
phase space derived from the K\"ahler potential can be seen to be precisely the 
metric on $S^2$. The emergence of $S^2$ for the Lie group $SU(2)$ is the key  
observation we want to record. For the groups $SU(3)$,$SU(4)$ we will construct 
appropriate K\"ahler forms which describe the classical phase space associated  with
these groups. It will then be demonstrated that the fibre obtained from the
orbit of the generators corresponding to simple roots of the Lie algebra
acting on this phase space    contain intersecting spheres.
 The spheres can easily be identified by the presence of nontrivial 
cycles with $\int \omega = 4\pi$ on the fibre and a sphere metric in an
appropriate subspace.The intersection properties of these spheres can  be
determined by using the methods of differential topology [6]. We demonstrate that
 the spheres described intersect in a manner precisely  mirroring the Dynkin
 diagram of the group. Such a result was established by Tits and Steinberg in a different setting (see the theorem by Tits and Steinberg in the article by Brieskorn E. in ref.[2]).We saw in section II that the spheres present when the     singular
 hypersurface considered there was resolved also contain intersecting spheres of exactly the same
 kind.  We now make use of the following theorems:\\

${\bf Chow's~Theorem~~ 3}~[6,7]$ $A$ $compact$ $hypersurface$ $can$ $always$ $be$
$represented$
$by$ $an$ $algebraic$  $variety$ $in$ $a$ $higher$ $dimensional$ $projective$ $space$.\\

${\bf Theorem~~ 4}~[7]$: $A$ $complex$ $curve$ $in$ ${\mathcal C}^n$ $can$ $always$ $be$
$embedded$ $in$ ${\mathcal C}^3$.\\

 The space constructed here is compact and hence the system can be represented 
as an algebraic variety and the intersecting spheres, present in ${\mathcal C
}^{n-1}$ , can be embedded in ${\mathcal C}^3$ .\\

With the help of these theorems we  see that the fibres containing the 
intersecting spheres can be embedded in ${\mathcal C}^3$. Thus the two
different mathematical objects: the resolved singular curve and the
algebraic curve present in the phase space bundle of $SU(n)$ both live in
${\mathcal C}^3$ and both contain the same number of spheres that intersect in  the
same way.

Now for the details. For $SU(3)$ we again work with the fundametal representation 
and introduce the coherent state
\begin{equation}
\vert {\nu_1},{\nu_2}\rangle = {e^{{{\nu}_1}{e_{13}}}}{e^{{{\nu}_2}{e_{23}}}}
\left(\matrix {0\cr 0\cr 1}\right) = \left(\matrix {{{\nu}_1}\cr {{\nu}_2}\cr 1}
\right)
\end{equation}
Note that $g=e^{{\nu_1}{e_{13}}+{\nu_2}{e_{23}}} = \left(\matrix {1 & 0 & \nu_1
\cr 0 & 1 & \nu_2\cr 0 & 0 & 1}\right)$ has generators $e_{13}$ , $e_{23}$ that
  commute . Furthermore $g$ commutes with $e_{13}$ and $e_{23}$ so that dimension
  of the centre of $g$ is 2 = rank of $SU(3)$. It is thus  
   a regular element in $SU(3,c)$ . The element is also unipotent as it is an
  upper triangular matrix with unit diagonal elements. 

The K\"ahler potential is given by
\begin{equation}
V(\nu_1,\nu_2,{\bar \nu_1},{\bar \nu_2}) \nonumber\\                       
 =  k\cdot \log{\langle {\nu_1},{\nu_2}\vert {\bar \nu_1},{\bar {\nu_2}}\rangle}
 \nonumber\\                                                                 
   =  k\cdot \log( 1+{\nu_1}{\bar{\nu_1}}+{\nu_2}{\bar \nu_2} ) \nonumber\\    
    =  k\cdot \log{\langle \nu \vert \nu \rangle}
\end{equation}
and the symplectic structure  determined by
\begin{equation}
{\omega_{i{\bar j}}} = k\cdot {\partial_{\nu_i}}{\partial_{\bar{\nu_j}}}V,\omega =
{\omega_{i{\bar j}}} d{\nu_i}\wedge d{\bar{\nu_j}} 
\end{equation}
is precisely that on ${\mathcal P}^2$ [6,8] in the coordinate chart ${{\nu^\prime}_0}
\ne 0$ , $({{\nu^\prime}_0},{{\nu^\prime}_1},{{\nu^\prime}_2})$ being the
 homogeneous coordinates and ${\nu_1} , {\nu_2} $ stand for ${\frac{{\nu^\prime
}_1}{{\nu^\prime}_0}}$ and ${\frac{{\nu^\prime}_2}{{\nu^\prime}_0}}$ . Also the
 symplectic structure (10) can be proved to be global [6] . It is then easy to 
verify that the commutation relations of $SU(3)$ are reflected in the Poisson 
brackets between the functions $\langle {e_{ij}}\rangle$ and $\langle{e_{kl}}\rangle$
where $\langle{e_{ij}}\rangle = {\frac{\langle{\nu_1}{\nu_2}\vert{e_{ij}}
\vert{\nu_1}{\nu_2}\rangle}{\langle{\nu_1}{\nu_2}\vert{\nu_1}{\nu_2}\rangle}}$\\
 
Note that
\begin{equation}  
[{\omega}] =  \left(\matrix {0 & a & 0 & b\cr -a & 0 & -{b^\prime} & 0\cr 0 & 
{ b^\prime} & 0 & c\cr -b & 0 & -c & 0}\right)                                   
\end{equation}

\begin{equation}
[{{\omega}^{-1}}] ={\frac{1}{ac-b{b^\prime}}}\left(\matrix {0 & -c & 0 & {b^\prime}
\cr c & 0 & -b & 0\cr 0 & b & 0 & -a\cr -{b^\prime} & 0 & a & 0}\right)
\end{equation}
where 
\begin{equation}
a = {\partial_{\nu_1}}{\partial_{\bar{\nu_1}}}\log{\langle \nu \vert \nu \rangle
}
 , b = {\partial_{\nu_1}}{\partial_{\bar{\nu_2}}}\log{\langle \nu \vert \nu \rangle}, 
      c = {\partial_{\nu_2}}{\partial_{\bar{\nu_2}}}\log{\langle \nu \vert \nu 
      \rangle} 
\end{equation}
and prime stands for complex conjugate .\\
We now proceed to construct fibres on this phase space:\\
The generators corresponding to the simple roots of $SU(3)$ are $e_{12}$ , $e_{23}$
and the group elements $\in {Z_+}$  are $e^{\mu e_{12}}$ ,  
  $e^{\lambda e_{23}}$. They are unipotent subregular elements $\in Z_+$ . To see
  this set $x=e^{\mu e_{12}}$ . Then $y = e^{\alpha e_{13} +\beta (h_1 +2h_2)+
 \gamma e_{32}+\delta e_{12}} \in SU(3,c)$ where $\alpha$, $\beta$, $\gamma$, $
\delta$ are complex parameters ,commutes with $x$. So dimension of the centre of
$x$  in $SU(3,c)$ is 4 = 2+2 = rank of $SU(3)$ + 2 . Thus $x$ is subregular in
$SU(3,c)$ . Similarly the element $e^{\lambda{e_{23}}}$ is subregular.We now 
consider the orbits of $e^{\mu e_{12}}$ and $e^{\lambda e_{23}}$ at the base point
on the phase space $\left(\matrix {\nu_1\cr \nu_2\cr 1}\right)$ .          
           
We have then $\mu$-orbit and $\lambda$-orbit as the fibre elements as 
$\left(\matrix {\nu_1 +\mu \nu_2\cr \nu_2\cr 1}\right)$ and $\left(\matrix 
{\nu_1\cr \nu_2 +\lambda\cr 1}\right)$. Relabelling the first orbit as 
$\left(\matrix {z_1\cr
 \nu_2\cr 1}\right)$ we can associate with it a K\"ahler potential $\log (1+ z_1 {
\bar z}_1)$. This is a sphere ${\mathcal P}^1$ . Similarly the $\lambda$ orbit 
is  also a sphere. To demonstrate that these two spheres intersect with intersection 
number 1 we consider ${e^{\mu e_{12}}}{e^{\lambda e_{23}}} \left(\matrix 
{\nu_1\cr \nu_2\cr 1}\right)$. The common $z_1$-$z_2$ (the first two coordinates) 
plane has an associated K\"ahler potential $\log(1+ z_1{{\bar z}_1}+
z_2{{\bar z}_2})$. The symplectic structure is then given by equation (10)
with $i,j=1 ,2$.  We now note the following theorems:\\

\noindent
{\bf {Theorem  5}}~[6] $The$ $de$ $Rham$ $cohomology$ $and$ $Dolbeault$ $cohomology$ $groups$ $for$
${\mathcal P}^n$ $are$ $related$: 
\[
{H_{\bar \partial}^{p,p}} ({{\mathcal P}^n}) \cong {H_{DR}^{2p}} \cong C
\]
{\bf {Theorem 6}}~[6] $The$ $intersection$ $of$ $the$ $two$ $surfaces$ ${\mathcal C}_i$ $and$
 ${\mathcal C}_j$ $are$ $given$ $by$ ${{\mathcal C}_i}\cdot{{\mathcal C}_j} = {\frac{
1}{(4\pi )^2}}\int {\omega_i}\wedge {\omega_j}$ $where$ $the$ $integration$ $is$ $in$ $the$
 $space$ $containing$ $the$ $surface$ ${\mathcal C}_i$ $and$ ${\mathcal C}_j$ $and$ $is$ $a$ $space$
 $of$ $dimension$ $four$ $and$ $\omega_i$ , $\omega_j$ $are$ ( $de$ $Rham$ $cohomology$ ) $elements$ 
 $of$ ${H^2_{DR}}(M,R)$.\\
\hspace{0.2cm} The cohomology groups associated with the symplectic form constructed are Dolbeault
cohomology groups while the intersection formula is valid for de Rham       cohomology groups.  
 However for ${\mathcal P}^n$ they are equivalent (theorem 5). We can thus
 determine intersection of spheres by simply evaluating ${\frac{1}{{(4\pi )}^2
}}\int \omega \wedge \omega$ , where the relevant four dimensional manifold (complex 
dimension 2 ) is  the common $z_1$-$z_2$ plane .  This is precisely seen 
to be one. Thus the two spheres on the fibre intersect once with intersection number one. 

The procedure outlined can be repeated for $SU(4)$ . This time
\begin{equation}
\vert \nu \rangle = \vert {\nu_1},{\nu_2},{\nu_3}\rangle = {e^{{\nu_1}{e_{14}}}
}{e^{{\nu_2}{e_{24}}}}{e^{{\nu_3}{e_{34}}}}\left(\matrix {0\cr 0\cr 0\cr 1}\right)
= \left(\matrix {{\nu_1}\cr {\nu_2}\cr {\nu_3}\cr 1}\right)
\end{equation}
and the K\"aler potential is
\begin{equation}
V({\nu_1},{\nu_2},{\nu_3},{\bar{\nu_1}},{\bar{\nu_2}},{\bar{\nu_3}}) = k\cdot 
\log{\langle \nu \vert  \nu\rangle} = k\cdot \log(1+{\nu_1}{\bar{\nu_1}}+
{\nu_2}{\bar{\nu_2}}+{\nu_3}{\bar{\nu_3}})
\end{equation}
Here again 
\begin{equation}
{\omega_{i,{\bar j}}} = {\partial_{\nu_i}}{\partial_{\bar \nu_j}} \log {\langle \nu \vert
 \nu \rangle}, \omega = {\omega_{i,{\bar j}}} d{\nu_i}\wedge d{\bar{\nu_j}} 
\end{equation}
is a symplectic structure on ${\mathcal P}^3$.The corresponding $[{\omega}]$ an
d $[{{\omega}^{-1}}]$ matrices are given by 
\begin{equation}
  [\omega ] = \left(\matrix {0 & a & 0 & b & 0 & c\cr -a & 0 & -{b^\prime} & 0 
& -{c^\prime} & 0\cr 0 & {b^\prime} & 0 & d & 0 & f\cr -{b^\prime} & 0 & -d & 0
 & -{f^\prime} & 0\cr 0 & {c^\prime} & 0 & {f^\prime} & 0 & g\cr -c & 0 & -f & 
0 & -g & 0\cr}\right) 
\end{equation}
where the primed entries stand for complex conjugates and

\vspace{0.5cm}

\begin{equation}
[{{\omega}^{-1}}] = {\frac{1}{N}}\left(\matrix {0 & dg-f{f^\prime} & 0 & f{c^\prime}
-g{b^\prime} & 0 & {b^\prime}{f^\prime}-d{c^\prime}\cr f{f^\prime}-dg & 0 
& bg-c{f^\prime} & 0 & cd-bf & 0\cr 0 & c{f^\prime}-bg & 0 & ag-c{c^\prime} & 0
 & b{c^\prime}-a{f^\prime}\cr g{b^\prime}-f{c^\prime} & 0 & c{c^\prime}-ag & 0 
& af-{b^\prime}c & 0\cr 0 & bf-cd & 0 & {b^\prime}c-af & 0 & ad-b{b^\prime}\cr 
d{c^\prime}-{b^\prime}{f^\prime} & 0 & a{f^\prime}-b{c^\prime} & 0 & b{b^\prime
}-ad & 0}\right)
\end{equation}

where 
\begin{eqnarray}
{N^2} = det {[\omega ]} , a = {\partial_{\nu_1}}{\partial_{\bar {\nu_1}}} \log {
\langle \nu \vert \nu \rangle}  , 
b = {\partial_{\nu_1}}{\partial_{\bar {\nu_2}}} \log {\langle \nu \vert \nu \rangle},
c = {\partial_{\nu_1}}{\partial_{\bar {\nu_3}}} \log {\langle \nu \vert
 \nu \rangle} \nonumber\\
d = {\partial_{\nu_2}}{\partial_{\bar {\nu_2}}} \log {\langle \nu \vert \nu \rangle},  f = {\partial_{\nu_2}}{\partial_{\bar {\nu_3}}} \log {\langle \nu \vert
 \nu \rangle} , 
g = {\partial_{\nu_3}}{\partial_{\bar {\nu_3}}} \log {\langle \nu \vert \nu \rangle}
\end{eqnarray}
and the prime denotes complex conjugate . Using $[{{\omega}^{-1}}]$ from eq.(18)
it is again straightforward to verify that the commutation relations of $SU(4)$
are reflected in the Poisson bracket between the functions $\langle {e_{ij}}
 \rangle$ and $\langle {e_{kl}} \rangle$ where again
$\langle {e_{ij}} \rangle \equiv {\frac{\langle \nu \vert {e_{ij}} \vert \nu 
\rangle}{\langle \nu \vert \nu \rangle}}$ . \\

Finally we look at the fibre . The unipotent subregular elements corresponding 
 to the simple roots are $e^{\mu e_{12}}$ , $e^{\lambda e_{23}}$ and $e^{\rho e_{34}}$.
 So we have $\mu$ , $\lambda$ and $\rho$ orbits on the fibre at the base
point $\left(\matrix {\nu_1\cr \nu_2\cr \nu_3\cr 1}\right)$. Each element acts
on a 2-complex dimensional subspace. It is easy to determine , as we have shown for  
the $SU(3)$ case that an individual orbit is ${\mathcal P}^1$ on the fibre.  Two 
generators intersect if they act on a common subspace i.e group elements
corresponding to $e_{12}$ and $e_{23}$ both act on the space labelled by 2,
while $e_{34}$ and $e_{12}$ have no common subspace. A differential topology way
of spotting this is to construct the vector obtained by acting $e^{\mu e_{12}}$
$e^{\lambda e_{23}}$ , say , on the base point. The variables $\mu$,$\lambda$
appear in the 1,2 position. Setting the variables corresponding to 1
,2 as $z_1$, $z_2$ the symplectic structure can be constructed from the 
corresponding K\"ahler potential and \\ ${\frac{1}{(4\pi )^2}}\int \omega \wedge
\omega  = {\frac{1}{(4\pi )^2}}\int   2(\omega_{z_1 {\bar z}_1}\omega_{z_2 
{\bar z}_2} - \omega_{z_1 {\bar z}_2}  
\omega_{z_2 {\bar z}_1}) dz_1 \wedge d{{\bar z}_1} \wedge dz_2 \wedge d{{\bar z }_2}$
gives the intersection. Similarly $e_{23}$ , $e_{34}$ will give rise to intersection
in the subspace $z_2$-$z_3$. The intersection corresponding to $e_{12}$, 
$e_{34}$ will be zero since they have no subspace in common. Here we have
 to evaluate $\int \omega \wedge \omega$ in the subspace 1,2 or 3,4 in each of 
 which $\omega \wedge \omega$ vanishes. Hence there is no intersection. 
 The intersection properties of the orbits described  easily generalises to the case 
 of $SU(n)$.For $SU(n)$ the symplectic structure will come from the group       elements constructed from the generators ${e_{12}}$, ${e_{13}}$,..${e_{1n}}$     while
 the simple roots are ${e_{12}}$,${e_{23}}$,..${e_{n,n-1}}$.
It is the orbit of the group elements generated by these simple roots acting
at any point on phase space that give fibres containing $n-1$ intersecting spheres.

\vspace{0.5cm}
\noindent
{\bf IV~~ Conclusions}\\
  We have shown in two examples , i.e for $SU(3)$ and $SU(4)$ how a classical
  phase space can be associated with these groups. In this phase space the orbit  
of the generators corresponding to the simple roots of the Lie algebra gives     
rise to intersecting spheres as fibres. 
 For $SU(n)$ the fibre of the bundle consists
of $n-1$ (equal to the rank of $SU(n)$) intersecting spheres. These spheres intersect precisely
as the negative of the Cartan matrix of $SU(n)$. A simple understanding of how   this happens 
is provided in our work. The structure of $SU(n)$, contained in the commutation
properties of the
simple
roots of its Lie algebra, is exactly the structure used to construct
the orbit of these generators in 
phase space. This structure in phase space gives rise to the resolved variety
associated with a singular variety.
 An algebraic group demonstration of the 
relationship between simple singularities of the ADE type and simply laced Lie 
groups of ADE type was proved by Brieskorn [2] where the role of the unipotent 
subregular elements of the groups was stressed and the earlier result regarding intersecting spheres of Tits and and Steinberg (  theorem by Tits and Steinberg discussed in the article by Brieskorn E. in ref.[2]) stated. Our explicit construction 
 uses unipotent regular elements of a special kind
 (viz.  ones involving mutually commuting generators) to construct classical
phase space and confirms the role played by subregular unipotent elements for 
making contact with resolved singularity. In our physically motivated approach  it is geometrically   very clear
why unipotent subregular elements are crucial: they are the group
elements that come from the simple roots of the Lie algebra. The classical
phase space for $SU(n)$ is shown explicitly to be contained in $G_c/B$ as
$CP^{n-1}$. The orbit of the generators of the simple roots of the
corresponding Lie algebra then provide a local trivialisation of a bundle
contained in $G_c/B$
with the classical phase space as the base. It is in the  fibre of this space
that  the variety corresponding to the resolved singularity is contained.
Our construction     extends easily to 
$SU(n)$. Extention of the construction described here to the D,E groups should be              straightforward.
It is pleasing that  classical phase space is where the resolved singular 
variety corresponding to ${x_0}{x_1}-{{x_2}^n}=0$ makes its appearence. 
 Our work thus provides  confirmation of the classical/quantum   correspondance discovered
in string theory between groups and singularities.

\vspace{0.2cm}

Acknowledgement: One of the authors(SKP) thanks Durga Paudyal(SNBNCBS,India),   
M.G.Mustafa(SINP,India) and Susanta Bhattacharya(RS College,Howrah,India) for   their unselfish help in some computer programmings.

\newpage
\noindent
{\bf Reference}\\

1. Aspinwall,Paul.S.: K3 Surfaces and String Duality; hep-th/961113
7. In S.T.Yau
(ed) $Differential$ $Geometry$ $inspired$ $by$ $String$ $Theory$. International Press 1999

2.  Brieskorn,E.: Singular Elements of Semi-Simple Algebraic 
Groups . $Nice$
$IM$ $Congress$,1970. Dimca,A.: $Singularities$ $and$
$Toplogy$ $of$ $Hypersurfaces$.
Springer-Verlag, 1992

3.Hull,C.,Townsend,P.K.:Unity of Superstring Dualities.Nucl. Phys. $\bf B438$,109-137 (1995).Harvey,J.,Strominger,A.:The heterotic String is a soliton.Nucl. Phys. $\bf B449$,535-552 (1995)

4. Perelomov,A.: $Generalized$ $Coherent$ $States$ $and$ $Their$                $Applications$. 
Springer, 1986

5. Pressley,A., Segal,G.: $Loop$ $Groups$ .Oxford Science Publication, 1986

6. Griffith,P.,Harris,J.: $Principles$ $of$ $Algebraic$ $Geometry$. 
John Wiley Sons,Inc.,Wiley Classics Library Edition,1974

7. Smith,K., Kahanpaa,L., Kekalainen,P.,Traven,W.: $An$ $Invitation$ $to$
$Algebraic$ 
$Geometry$.Springer,2000

8. Arnold,V.I.: $Mathematical$ $Methods$ $of$ $Classical$ $Mechanics$.
Springer-Verlag,1978

\end{document}